\documentclass{apt} 

\renewcommand{\authornames}{Szapudi \& Szalay} 
\renewcommand{\shorttitle}{$N$-point Estimators} 

\newcommand{\tw}{\hat{w}}
\newcommand{\tK}{\hat{K}}
\newcommand{\avg}[1]{\left\langle{#1}\right\rangle}

\begin{document}

\title{ The Variance of a New Class of $N$-point Correlation Estimators in
Poisson and Binomial Point Processes }

\authorone[Durham University]{Istv\'an Szapudi}

\authortwo[Johns Hopkins University]{Alexander S. Szalay}

\addressone{University of Durham, Dept. of Physics, South Rd, Durham, DH1 3LE,
 England }

\addresstwo{Johns Hopkins University, Dept. of Physics and Astronomy, 
  Baltimore, MD 21218 }

\begin{abstract}
We describe a set of new estimators for the N-point
correlation functions of point processes. The variance of these estimators
is calculated for the Poisson and binomial cases.
It is shown that the variance of the unbiased estimator
converges to the continuum value much faster
than with any previously used alternative, all terms with slower
convergence exactly cancel. We compare our estimators
with Ripley's $\tK_0$ and $\tK_2$.

\end{abstract} 


\ams{}{} 

\section{Introduction}

Estimation of correlation functions from a set of
points is a classical problem of spatial statistics.
The two-point correlation functions are the most widely used, but
there is an increasing interest in estimating higher order 
correlation functions as well.
A new class of estimators was introduced in astrophysics \cite{ss98} 
which pertains to most methods currently applied to
data sets of galaxy positions. 
We present the rigorous calculation of the variances of
these new estimators for Poisson and binomial point
processes. It will be shown in the next sections that
for these processes the estimators have smaller variance
than any estimator previously used for the same statistics.
While the results are completely general, they were motivated
by future astrophysical applications, therefore all the examples
will be taken from there. Higher order correlation
functions from galaxy catalogs are routinely estimated since
the 70's. With a new generation of galaxy catalogs coming
on line in the next five years, understanding the estimators
is an important and timely problem.

A unique feature of spatial 
statistics is that errors of a measurement are often
dominated by geometrical terms, like edge effects \cite{ripley88}.
Thus ever since correlation functions were estimated, 
corrections for edge effects have played a central role.
The estimators for correlation functions
in the astrophysical literature is reviewed in \cite{ss98},
here we only quote a few selected additional references
\cite{hew82,dp83,ls93,h93}. It is generally accepted
that the most efficient estimator
for the two-point function 
is that of \cite{ls93}, or its relative
\cite{h93}. 

The new estimator can be approximated in the 
continuum limit (achieved when the number density of
data points $\rightarrow\infty$) as $\tw_2 = \avg{\delta_1 \delta_2}$,
where $\avg{}$ denotes ensemble average, and
$\delta$ is the fluctuation of the continuous
(galaxy) density field $\rho$. It is defined as
$\rho = \avg{\rho}(1+\delta)$, thus $\avg{\delta} = 0$.
The Monte Carlo representation of this 
estimator is often written symbolically as
$(DD-2 DR+RR)/RR$, with $DD$, $DR$, and $RR$ representing the
respective pair counts. The important point is that
the above estimator contains only the most necessary terms
in the continuous limit, 
while all others, such as  
$DD/RR-1 \rightarrow \avg{(1+\delta_1)(1+\delta_2)}-1 = 
\avg{\delta_1\delta_2+\delta_1+\delta_2}$,
have extra terms. These extra terms do not affect
the ensemble average of the unbiased estimator, 
but increase the variance. 
The deceptively simple look of the second estimator in terms of $D$ and $R$
was the reason for its popularity.

Ripley \cite{ripley88} has discussed extensively the variance of second
order estimators for Poisson and binomial point processes. 
He has shown, that the ${\cal O}(n^{-1})$term 
in the variance of the simple estimator, $\hat K_0$, 
is proportional to $u$, the perimeter for a
two dimensional domain. This implies that the effect is due
to inadequate edge corrections, in agreement with Hewett's \cite{hew82}
suggestion. The subtraction of the appropriate $DR$ terms is
equivalent to an edge correction.

The effect of the extra terms on the variance
is even more pronounced for the higher order functions, since there
will be a lot more terms arising through various combinatorial
expressions. We proposed intuitively \cite{ss98} 
that the obvious generalization for
higher order correlations is to create  
higher order equivalents of the estimator \cite{ls93}.
With $\delta_i \simeq (D_i-R_i)/R_i$
this corresponds to $\avg{\delta_1...\delta_N}$.  In
symbolic notation, this estimator can be written as
 \begin{equation} 
	\tw_N = {(D_1-R_1).(D_2-R_2)\ldots (D_N-R_N)}/R_1\ldots R_N.
        \label{eq:est}
  \end{equation} 
This corresponds to the Monte-Carlo approximation of 
the continuum limit of the $N$-point
correlation function of {\em fluctuations};
the exact meaning is discussed later. The most significant
result is that the correlations
of the fluctuations automatically correct for edge effects
for Poisson and binomial point processes.

Different approaches
to edge effects exist; for a review see \cite{stst98,k98}.
It remains to be seen whether these geometrically motivated estimators,
or their generalization for higher order \cite{han83}
fare better than the \cite{ls93} estimator, or the related estimator
introduced by \cite{du89}. 

The main goal of this article is to rigorously derive the
variance of the estimator for Poisson
and binomial processes. 
The next section presents the analytic calculation of
the variance for arbitrary $N$.
Section 3 compares the second
order estimator with that of the related Ripley's $K$ function.
The last section summarizes the results.

\section{Variance of the Edge Corrected Estimators in the Poisson
and Binomial Point Processes}

Many interesting statistics, such as the
$N$-point correlation functions and their Fourier analogs, can be
formulated as functions over $N$ points from the catalog. The
covariance of a pair of such estimators will be calculated for
Poisson and binomial point processes.  They correspond to the cases, where
the number of detected objects is varied or fixed a priori.  
The general case, where correlations are non-negligible is
left for future work.
The following calculations heavily rely on the elegant
formalism outlined in Ripley \cite{ripley88}, which can be consulted for
details.

Let $D$ be a catalog of data points to be analyzed, and $R$ randomly
generated over the same area, with averages $\lambda$, and $\rho$
respectively.  The role of $R$ is to perform a Monte Carlo integration
compensating for edge effects, therefore eventually the limit $\rho
\rightarrow \infty$ will be taken. $\lambda$ on the other hand is
assumed to be externally estimated with arbitrary precision. We also
assume that the correlations in the point process are weak, i.e. we
operate in the Poisson limit.

Let us define symbolically an estimator $D^pR^q$, with $p+q = N$ for a 
function $\Phi$ symmetric in its arguments
  \begin{equation}
	D^pR^q = \sum\Phi(x_1,\ldots,x_p,y_1,\ldots,y_q),
  \end{equation}
with $x_i \ne x_j \in D, y_i \ne y_j \in R$. For example for the
two point correlation function $\Phi(x,y) = [x,y \in D, r \le d(x,y) \le
r+dr]$, where $d(x,y)$ is the distance between the two points, and
$[condition]$ equals $1$ when $condition$ holds, $0$ otherwise.
Ensemble averages can be estimated via factorial moment
measures, $\nu_s$ \cite{dv72,ripley88}. 
In the Poisson limit $\nu_s = \lambda^s\mu_{s}$, 
where $\mu_s$ is the $s$ dimensional Lebesgue measure.

The general covariance of a pair of estimators is
  \begin{equation}
	\avg{D_a^{p_1}R_a^{q_1}D_b^{p_2}R_b^{q_2}} = 
	\sum_{i,j}{p_1 \choose i}{p_2 \choose i} i!
	{q_1 \choose j}{q_2 \choose j} j!\,
	S_{i+j}\lambda^{p_1+p_2-i}\rho^{q_1+q_2-j},
  \label{eq:cov}
  \end{equation}
with $p_1+q_1 = p_2 + q_2 = N$, and 
  \begin{equation}
	S_{k} = \int\Phi_a(x_1\ldots x_{k},y_{k+1}\ldots y_N)
              \Phi_b(x_1\ldots x_{k},z_{k+1}\ldots z_N)
              \mu_{2N-k}.
  \end{equation} 
Throughout the paper we use the convention that ${k\choose l}$ is nonzero
only for $k \ge 0, l\ge 0$, and $ k \ge l$. 
Here $\Phi_a$ and $\Phi_b$ denote two different functions, for instance
corresponding to two radial bins.  The
expression simply describes the fact that out of the $p_1$ and $p_2$
different data points in $D$ we have an $i$-fold degeneracy, as well
as a $j$-fold degeneracy in the random points drawn from $R$. For each
of these configurations the geometric phase-space $S_{i+j}$ is
different, and we sum their contributions. The dependence of
$S_k$ on $a,b$, and $N$ is not noted for convenience, but they will be
assumed throughout the paper.  An estimator for the generalized
$N$-point correlation function is
  \begin{equation}
  	\tw_N = \frac{1}{S}  \sum_i 
	{N \choose i}(-)^{N-i} (\frac{D}{\lambda})^i (\frac{R}{\rho})^{N-i},
  \end{equation}
where $S = \int\Phi\mu_{N}$ (without subscript).  This definition can
be expressed as $(\hat D-\hat R)^N$, where $\hat{\,} $ means
normalization with $\lambda, \rho$ respectively.  In this symbolic
$N$th power, each factor is evaluated at a different point.  Simple
calculation in the limit of zero correlations yields $\avg{\tw_N} =
0$.
\begin{theorem} 
The asymptotic covariance between two estimators of the above form for a
Poisson point process in the limit of $\rho\rightarrow\infty$ is
\begin{equation}
   {\rm (co)Var_\lambda}\, 
   \tw_N =  \avg{\tw_{a,N}\tw_{b,N}} = \frac{S_N N!}{S^2\lambda^N}.
   \label{eq:th1}
\end{equation}
\end{theorem}

{\it Proof.} According to Eq. \ref{eq:cov} the covariance can be written
as
\begin{equation}
  \avg{\tw_{a,N}\tw_{b,N}} = \sum_{i_1,i_2,i,j}{N \choose i_1}{N \choose i_2}
  {i_1 \choose i} {i_2 \choose i}i! {N - i_1 \choose j}{N - i_2 \choose j}j!
  \frac{S_{i+j}}{S^2}\lambda^{-i}\rho^{-j} (-)^{2N-i_1-i_2}.
\end{equation}
In the interesting limit, where $\rho \rightarrow \infty$ only $j = 0$ 
survives. Changing the order of summation yields
\begin{equation}
   \avg{\tw_{a,N}\tw_{b,N}}= \frac{1}{S^2}\sum_i S_i \lambda^{-i} i! f_{Ni}^2,
\end{equation}
with
\begin{equation}
  f_{Ni} = \sum_j {N \choose j}{j \choose i} (-1)^{N-j}.
\end{equation}
This latter can be identified as the coefficients of $\sum_N(xy)^N$,
therefore $f_{Ni} = \delta_{Ni}$. This in
turn proves the theorem noting that $\avg{\tw_N} = 0$.
This formula represents both variance and covariance
depending on whether in the definition of $S_N$ the implicit
indices $a$ and $b$ are equal or not. 

While in the Poisson model the total number of points in the
domain can vary, it is fixed in the binomial model. This latter
case corresponds to surveys, that detect a certain number of galaxies,
and use that to estimate the mean density as well. In a sense, this would
be the conditional estimator of the correlations given the number of
galaxies. The normalization of the estimator
changes slightly: $\lambda^i \rightarrow (n)_i/v^i$, 
where $n$ is the total number of objects in the survey, 
and $(n)_i = n(n-1)\ldots (n-i+1)$ is the $i$-th falling factorial.
This renders the definition of the estimator for binomial process as
  \begin{equation}
  	\tw_N = \frac{1}{S}  \sum_i 
	{N \choose i}(-)^{N-i} \frac{(D v)^i}{(n)_i}\frac{(R v)^{N-i}}
        {(n_r)_{N-i}},
  \end{equation}
where $n_r$ is the number of points in the auxiliary random process
$R$.

For a binomial process the
factorial moment measure is $(n)_N v^{-N} \mu_N$, with $v$, the volume
of the survey.
This fact enables the proof of the following theorem. 

\begin{theorem}
The asymptotic covariance of two estimators for a binomial point process 
in the limit of $n_r\rightarrow\infty$ is
  \begin{equation}
  	{\rm (co)Var_n}\, \tw_N  = \frac{1}{S^2}
        {n \choose N}^{-1}\sum_i S_i v^i {N \choose i} (-)^{N-i}.
  \end{equation}
\end{theorem}

{\it Proof.} First it is convenient to prove the following lemma

\begin{lemma}
For all possible integer values of $N,n$,
and $i$
  \begin{equation}
    \sum_{i_1,i_2}{n-i_2\choose N - i_2}{n - i_1 \choose i_2 - i}
         {N - i \choose i_1 - i}(-)^{i_1+i_2} = (-)^{N-i},
  \end{equation}
where the summation is over all possible values of $i_1$ and $i_2$.

\end{lemma}

{\it Proof.} It follows by
induction over $N$. For $N = i$ it is true, since 
$N = i = i_1 = i_2$ are the only possible values. Thus
  \begin{equation}
    {n-i\choose 0 }{n-i\choose 0 }
    {0\choose 0 } (-)^{2i} = (-)^0
  \end{equation}
for any $n$ and $i$. Assume it is true for a particular $N$ for
any $n$ and $i$. Then for $N+1$
  \begin{equation}
    \sum_{i_1,i_2}{n-i_2\choose N+1 - i_2}{n - i_1 \choose i_2 - i}
         {N+1 - i \choose i_1 - i}(-)^{i_1+i_2} = (-)^{N+1-i}. 
  \end{equation}
  By introducing $m = n-1, k = i -1$, $k_1 = i_1 -1$, and
  $k_2 = i_2 -1$ this reads
  \begin{equation}
    \sum_{k_1,k_2}{m-k_2\choose N - k_2}
     {n - k_1 \choose k_2 - k}
         {N - k \choose k_1 - k}
         (-)^{k_1+k_2} = (-)^{N-k},
  \end{equation}
  which is true by induction.

Now to prove the theorem consider the equation for the covariance using
the appropriate factorial moment measure for binomial point process.
After $n_r\rightarrow\infty$
\begin{equation}
 \avg{\tw_{a,N}\tw_{b,N}}  = \frac{1}{S^2}\sum_{i,i_1,i_2}S_i v^i
              i!{N \choose i_1}{N \choose i_2}
             {i_1 \choose i}{i_2 \choose i}
             \frac{(n)_{i_1+i_2-i}(-)^{i_1+i_2}}{(n)_{i_1}(n)_{i_2}}.
\end{equation}
Applying the lemma for each $i$ separately, the theorem is proven.
  \begin{eqnarray}
    &&\sum_{i_1,i_2} {N\choose i_1} {N\choose i_2}
                   {i_1 \choose i} {i_2\choose i} (-)^{i_1+i_2}
                   (n-i_1)_{N-i_1}(n-i_2)_{N-i_2}(n)_{i_1+i_2-i} i!
   = \nonumber \\
   &&N! (n)_N {N \choose i}(-)^{N-i}.
  \end{eqnarray}

For $N = 2$ the theorem coincides with \cite{ls93}, taking into account
that $S^2 = S_0 \simeq (S_2 v^2)^2$.

\section{Discussions}

For practical applications the function $\Phi$ has to be
specified. For instance, $\Phi = 1$ when the
$N$-tuplet satisfies a certain geometry (with a suitable bin
width), and $0$ otherwise 
yields the total (or disconnected) $N$-point correlations of the fluctuations
of the process. See \cite{ss98} for detailed discussion of
possible choices for the function $\Phi$ to render popular
statistics for the distribution of galaxies, such as
power spectra, cumulant correlators, etc. Here we concentrate
on the comparison with other second order estimators.

The number of neighbors from a point within a distance of $\le t$
is defined as $\lambda K(t)$ \cite{ripley88}.
A family of estimators denoted by
$\tK_i$  was introduced \cite{ripley88} with subtle differences in
edge correction. It was found that Ripley's $\tK_2$ has the smallest
variance of all. Similar conclusions were reached in
a numerical setting motivated by potential astrophysical 
applications \cite{k98}. The difference between these
estimators and ours is twofold: first, the normalization is different,
second, they estimate the moments of the full point process
while $\tw_N$ deals with the moments of the {\em fluctuations}.
The former point is trivial, while the latter is crucial, as
shown later. When $\tw_2$ is used to extract the (differential)
two-point correlation function, the connection with 
Ripley's cumulative $K$ can be expressed as
$K(t) = \lambda \int_0^t dr (1+w_2(r))$.
Note that in astrophysics the prevailing choice {\em is} $w$,
and $K$ for the most part would be estimated for the purpose
of eventually obtaining $w$ from it. 
Nevertheless, in other
disciplines, or perhaps even for certain aspects of astrophysics,
$K$ could be more advantageous. 

Next we show that $\tw_2$ has smaller
variance than any of the $\hat K_i$. 
For a pair-estimator $\hat{T} = \sum_{x\ne y}\Phi(x,y)$ the variance
for a Poisson process can be expressed in a quite general fashion.
Using the notation of the previous section, the variance is
${\rm Var}_\lambda \hat{T}= 4\lambda^3 S_1+2\lambda^2S_2$ \cite{ripley88}.
Ripley has derived an approximation of this formula
for the ``naive'' estimator $\tK_0(t) = aT/n^2$,
$   {\rm Var}_\lambda \tK_0(t) = 
   \frac{1}{\lambda^2}\left[\frac{\pi t^2}{a}-2\frac{u t^3}{a^2}
   +1.34\lambda\frac{ut^5}{a^2}\right],
$
where $a$ is the area of the two-dimensional domain, and
$u$ is its perimeter.

In order to compare the variance $\tw_2$ and $\hat{T}$
($\tK_0$ up to normalization), we scale $\hat{T}$ by
$S\lambda^{2}$
  \begin{eqnarray}
    {\rm Var_\lambda}\, \tw_2  = 
    &&\frac{2 S_2}{S^2\lambda^2},  \nonumber \\
    {\rm Var}_\lambda\left(\frac{\hat{T}}{S\lambda^{2}}\right) = 
    &&\frac{2 S_2}{S^2\lambda^2} +
    \frac{4 S_1}{S^2\lambda}.  
    \label{eq:varcompare}
  \end{eqnarray}
The Poisson terms in the number of pairs, 
$\simeq\lambda^{-2}$, are identical, while the
${\cal O}(1/\lambda)$ terms are missing from the variance of
$\tw_2$. The latter can be appreciable when $\lambda$ is small
for  the estimator $\tK_0$ based on $T$.

The most clever estimators of $K$, 
such as Ripley's $\tK_2$, suppress this term
considerably, nevertheless it is always present.
E.g., the variance of $\tK_2$ for a Poisson process is \cite{ripley88} 
$   {\rm Var}_\lambda \tK_2(t) = 
   \frac{1}{\lambda^2}\left[\frac{\pi t^2}{a}+0.96\frac{u t^3}{a^2}
   +0.13\lambda\frac{ut^5}{a^2}\right],
$

In general for any $N$, all contributions 
${\cal O}(\lambda^{-k}$), $k < N$, exactly cancel for
$\tw_N$.  
Since terms higher than ${\cal O}(\lambda^{-N})$ are absent, the
asymptotic behavior of our estimator is optimal: the
only possible improvement for Poisson (or binomial) processes
is perhaps to decrease the multiplicative factor.

In fact $\tK_2$ suppresses the $\lambda^{-1}$ term
at the expense of boosting the coefficient
of $\lambda^{-2}$ compared to $\tK_0$. 
In contrast, for $\tw_2$ no such boost
is present according to Equation~\ref{eq:varcompare}, 
the coefficients of 
the Poisson term are identical to that of the $\tK_0$ estimator. 
This suggests that the estimator $\tw_N$ is probably close
to optimal for Poisson (and binomial) processes. We conjecture
that this is approximately true for many correlated point 
processes as well, although for extreme cases, such 
a line segment process, a small bias was noted by the
numerical investigations of \cite{k98} for both $\tw_2$, and
for other estimators related to the $K$ function. 

The intuitive meaning of the results for binomial process
is clear from Theorem 2.:
the variance of the estimators is $\propto {n\choose N}^{-1}$, i.e.
Poisson in the number of $N$-tuplets, with multiplicative
factors depending on the available geometric phase space.
For the Poisson process, Theorem 1. is
identical up to discreteness effects: it is
inversely proportional to the probability
density of finding $N$-points in the domain,  $\propto(\lambda^N)^{-1}
\simeq (n/v)^{-N}$.

Our approach to  suppressing the offending non-Poisson terms
involves an auxiliary Poisson
process ($R$), for which the average density is assumed
to be infinity $\rho\rightarrow\infty$. The limit is
not possible in practice, therefore it is
desirable to evaluate the speed of convergence. 
A calculation analogous to the proof of Theorem 1. shows
that
\begin{equation}
   {\rm (co)Var_{\lambda,\rho}}\, \tw_N = \frac{S_N N!}{S^2\lambda^N}
   \left\{ 1+\frac{1}{N}\left(\frac{\lambda}{\rho}\right)+
   {\cal O}(\lambda/\rho)^2 \right\}. 
   \label{eq:th1plus} 
\end{equation}
This generalization of Theorem 1. clearly shows that for a finite
auxiliary Poisson process $R$, there is indeed a term with $1/\lambda$,
but suppressed by a large factor. 
For instance, the relative  non-Poisson correction
is $\lambda/(2\rho)$ for $N= 2$, giving less than a percent contamination
when the auxiliary process is more than fifty times the original
process to be measured. This is fairly convenient,
since, unlike for the original process,
we have full control over the artificially introduced process.
Note that the above considerations assume exact knowledge
of the average count $\lambda$, thus no conditioning
on the number of points is assumed. 

For a binomial process the argument is exactly analogous to
the previous one, therefore we only outline it briefly.
The variance for a general estimator $T$ is \cite{ripley88}
${\rm Var}_nT = 4n(n-1)(n-2)a^3S_1-n(n-1)(4n-6)a^{-4}S^2+
          2n(n-1)a^{-2}S_2$.
This is to be compared with Theorem 2.
${\rm Var_n}\, \tw_2 = 2[S_0-2S_1a+S_2a^2]/S^2n(n-1)$ (replacing $v$ with
$a$ for two dimensions). Again $T (n(n-1))^{-2}a^4$ has
to be considered because of the normalization. The first two
terms in the previous equation will yield $n^{-1}$ terms,
while the variance of $\tw_2$ is again Poisson in terms of the
number of pairs. Finally, the convergence properties are
expected to be similar to the Poisson process, although
we have not performed the calculations. 

It is worth to emphasize again that, while both are two-point measures,
$w$ and $K$ are slightly different objects. In astrophysics
$w$ is the desired quantity, but in some cases $K$ might be
more advantageous; then Ripley's $\tK_2$ is
still the preferred estimator.
Clearly, these findings remain true for $N > 2$ as well. 

\section{Summary}

In summary we have calculated the variance of a new class
of estimators $\tw_N$ for the $N$-point correlation functions
for Poisson and binomial point processes. 
The results were compared with variances concerning a different class of
estimators based on the $K$ function. The main difference
is that $\tw_N$ estimates the $N$-point correlation function
of the fluctuations of the point process, while $\tK_i$ estimate the
(cumulative) moments of the full point process. This property
apparently renders the edge effect correction in $\tw_N$ exact, 
leaving only the Poissonian contribution to the variance in terms
of $N$-tuplets, i.e. terms $\propto 1/\lambda^N$, or $1/{n\choose N}$ 
for Poisson and binomial processes, respectively. All lower order
terms, dominating for sparse processes, exactly cancel.
This is not true for any of the estimators for $K$, although the best
ones achieve a significant suppression of the offending
non-Poisson terms at the price of boosting the constant factor
multiplying the Poisson terms with respect to $\tK_0$, the naive
estimator. The speed of convergence was 
calculated for the new estimator $\tw_N$, which assumed an auxiliary
Poisson process $R$, with the average $\rho\rightarrow\infty$.
The leading order non-Poisson term in the variance was found
to be suppressed by a large factor $\lambda/\rho N$, thus the
convergence is controllable in practice when 
$\rho$ is finite.

I.S. was supported by a PPARC rolling grant for Extragalactic Astronomy
and Cosmology at Durham. A.S. was supported by NASA and the NSF.
We would like to thank Martin Kerscher for useful comments.

\hfill\newpage

\def\apj{ApJ}
\def\apjl{ApJL}
\def\apjs{ApJS}
\def\aap{A\&A}
\def\mnras{MNRAS}

{}

\end{document}